\documentclass[final,superscriptaddress,english,twocolumn,amssymb,aps,prx,
longbibliography]{revtex4-1}
\usepackage{graphicx}
\usepackage{natbib}
\usepackage{amsmath}
\usepackage{amssymb}
\usepackage{appendix}
\usepackage{soul}
\usepackage[dvipsnames]{xcolor}{\huge }
\usepackage[section]{placeins}
\definecolor{darkblue}{rgb}{0,0,.65}
\definecolor{darkgreen}{rgb}{0.28,0.41,0.19}

\usepackage{nicefrac}
\usepackage[%
    pdfauthor={David J. Luitz},%
  pdfstartview=FitH,%
  breaklinks=true,%
  bookmarks=true,%
  colorlinks=true,%
  anchorcolor=black,%
  citecolor=blue,
  filecolor=black,%
  menucolor=black,%
  urlcolor=darkblue,%
  linkcolor=blue,%
 ]{hyperref}
\usepackage[all]{hypcap} %
\newcommand{\bra}[1]{\langle\,#1\,|}
\newcommand{\ket}[1]{|#1\rangle}

\graphicspath{{images/}}
\begin{document}

\title{Possible inversion symmetry breaking in the $S=1/2$ pyrochlore Heisenberg magnet}

\author{Imre Hagym\'asi}\email{hagymasi@pks.mpg.de}
\affiliation{Max Planck Institute for the Physics of Complex Systems, 
Noethnitzer Str. 38, 01187 Dresden, Germany}
\affiliation{Strongly Correlated Systems "Lend\"ulet" Research Group, Institute 
for Solid State
Physics and Optics, Wigner Research Centre for Physics, Budapest H-1525 P.O. 
Box 49, Hungary
}
\author{Robin Sch\"afer}\email{schaefer@pks.mpg.de}
\affiliation{Max Planck Institute for the Physics of Complex Systems, 
Noethnitzer Strasse 38, 01187 Dresden, Germany}
\author{Roderich Moessner}\email{moessner@pks.mpg.de}
\author{David J. Luitz}
\email{dluitz@pks.mpg.de}
\affiliation{Max Planck Institute for the Physics of Complex Systems, Noethnitzer Str. 38, 01187 Dresden, Germany}

\date{\today}

\begin{abstract}
	We address the ground-state properties of the long-standing and much-studied three-dimensional quantum spin liquid candidate, the $S=\frac 1 2$ pyrochlore Heisenberg
	antiferromagnet. By using $SU(2)$ DMRG, we are able to access cluster sizes of up to 128 spins. Our most striking finding is a robust spontaneous inversion symmetry breaking, reflected in an energy density difference between the two sublattices of  tetrahedra, familiar as a starting point of earlier perturbative treatments. We also determine the ground-state energy, $E_0/N_\text{sites} = -0.490(6) J$, by combining extrapolations of DMRG with those of a numerical linked cluster expansion. These findings suggest a scenario in which a finite-temperature spin liquid regime gives way to a symmetry-broken state at  low temperatures.
\end{abstract} 

\maketitle

\begin{figure}[h]
	\includegraphics[width=\columnwidth]{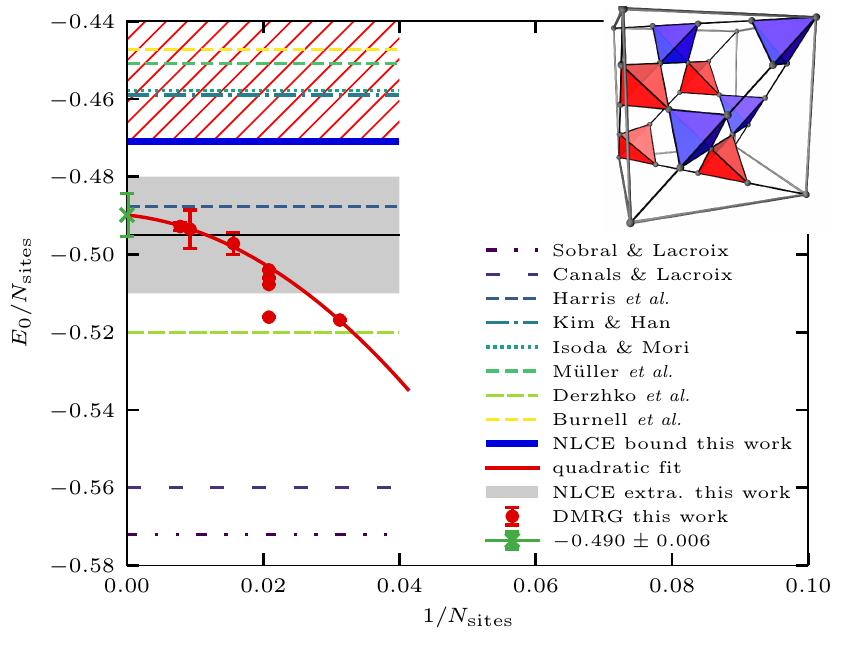}
	\caption{Ground-state energies from various 
		approaches. The horizontal lines denote the predictions for 
		the ground-state energy per site ($J=1$) in the 
		thermodynamic limit: Sobral and Lacroix $-0.572$ [\onlinecite{sobral_1998}], 
		Canals and Lacroix $-0.56$  [\onlinecite{canals_lacroix_prb_2000}], Derzhko 
		\emph{et al.} $-0.52$ [\onlinecite{derzhko_adapting_2020}], Harris \emph{et al.} 
		$-0.487$ [\onlinecite{harris_ordering_1991,koga_prb_2001}], Kim and Han 
		$-0.459$,  [\onlinecite{kim_prb_2008}], Isoda and Mori $-0.4578$ 
		[\onlinecite{isoda_valence_bond_1998}], M\"uller \emph{et al.} $-0.4509$
		[\onlinecite{muller_thermodynamics_2019}], Burnell  \emph{et al.} $-0.4473$
		[\onlinecite{burnell_monopole_2009}].  
		The solid red points are our DMRG results for periodic clusters, extrapolated 
		to infinite bond dimension using a quadratic polynomial. 
		The thick blue line represents a robust upper bound for the ground-state energy, 
		obtained from converged NLCE results at finite temperature, thus excluding the red 
		hashed area.
		The solid black line shows the extrapolated value of the converged NLCE 
		results to zero temperature (\textit{cf.} Appendix \ref{sec:nlce}), and the 
		gray shaded area indicates the confidence interval of this extrapolation. The inset shows the cubic unit cell of the pyrochlore lattice, highlighting the two tetrahedral sublattices in red and blue.
	}
	\label{fig:energy_comparison}
\end{figure}

\emph{Introduction.---} Frustrated magnets, on account of exhibiting many competing low energy states, are a fertile ground for exotic physics. 
A celebrated example is the pyrochlore Heisenberg antiferromagnet, which 
resides on a lattice of corner sharing tetrahedra, depicted in the
inset of Fig.~\ref{fig:energy_comparison}.
The classical Heisenberg model on this lattice has a highly degenerate ground state \cite{villain_insulating_1979}, 
forming a classical spin liquid \cite{moecha_pyro_prl} with an emergent gauge field \cite{Isakov_dipolar_prl}.

In contrast, the ground state of the quantum pyrochlore antiferromagnet remains enigmatic. 
While recent experimental evidence in the approximately isotropic $S=1$ compound NaCaNi$_2$F$_7$ shows a liquid like state down to low temperature \cite{plumb_continuum_2019}, the $S=1/2$ case is still open both in theory and experiment.

Theory work on this prominent quantum spin liquid candidate over the years has been formidable.
Absent a systematically controlled method, various approaches have somewhat inevitably led to 
an array of possible scenarios. One strand of work has built on a 
perturbative approach, in which half the couplings (those on one tetrahedral sublattice) are switched on perturbatively. 
This has led to suggestions of a ground state which breaks translational and rotational symmetries \cite{harris_ordering_1991,tsunetsugu_prb_2001,Tsunetsugu_pyro_2001}, a valence bond crystal \cite{isoda_valence_bond_1998} or a spin liquid state \cite{CanalsLacroix_prl}. 
On top of this,  the contractor renormalization method \cite{Berg_subcontractor_2003} finds antiferromagnetic ordering in a space of supertetrahedral pseudospins, pointing to an even larger real-space unit cell. To render the problem more tractable,  
all these theories involve the derivation of an effective Hamiltonian, which is \textit{per se} not exactly solvable and hence solved by some type of approximation, ranging from mean field theory to classical Monte Carlo numerics. %
On a different axis in theory space, parton-based theories  yield an ordered state with a chiral order parameter \cite{kim_prb_2008} or a monopole flux state \cite{burnell_monopole_2009}, while the pseudofermion functional renormalization group suggests a spin liquid ground state \cite{iqbal_quantum_2019}. 

In view of this relatively wide range  of ground-state candidates, a controlled 
and unbiased treatment of the model is clearly desirable, if only to narrow the 
possible location of the goalposts somewhat. Unfortunately, most numerical approaches quickly reach 
their limits for frustrated magnets in $d=3$. While exact diagonalization is 
currently limited to $\sim 48$ sites \cite{lauchli_kagome_2019}, possible 
alternatives are series expansions such as the numerical linked cluster 
expansion (NLCE) 
\cite{rigol_nlce_kagome_2006, rigol_nlce_square_2007, rigol_nlce_kagome_square_tri_2007, 
	khatami_nlce_checkerboard_2011, khatami_nlce_clinoatacamite_2011, 
	khatami_nlce_pinwheel_kagome_2011,  khatami_nlce_optical_lattic_2011, khatami_nlce_hubbard_2011, applegate_nlce_pyrochlore_exp_comp_2012, 
	singh_nlce_tetra_pyrochlore_2012, tang_nlce_2013, hayre_nlce_pyrochlore_2013, 
	jaubert_nlce_pyrochlore_exp_comp_2015, benton_nlce_pyrochlore_qsl_2018, 
	benton_nlce_pyrochlore_2018, pardini_nlce_pyrochlore_qsl_2019,  
	schafer_pyrochlore_2020} or high temperature expansions 
\cite{lohmann_tenth-order_2014,richter_combining_2019}, which can be pushed down to low temperatures \cite{schafer_pyrochlore_2020}, 
although they do not provide access to the ground state itself and are particularly challenged by many competing low energy states.

To access the ground-state wave function directly, the DMRG method --- originally 
devised in one dimension 
\cite{white_1992,white_1993,noack2005,schollwock_review_2011,hallberg_review} has been pushed to 
two dimensions, in particular for the two-dimensional cousin of pyrochlore, the 
kagome antiferromagnet 
~\cite{schollwock_prl_2012,Jiang2012,yan_science_2011,pollmann_prx_2017, 
jiang_prl_2008}. 
 
 Here, we take DMRG one step further, by applying it to the pyrochlore lattice in $d=3$, 
and present a  study of  periodic clusters with 
$N_\text{sites}=32, 48, 64, 108, 128$. This demonstrates that DMRG can treat 
clusters with up to $128$ sites reliably, significantly larger than previous 
exact diagonalization results of 36 sites \cite{chandra_spin12_2018}. 
Exploiting the SU(2) symmetry of the model \cite{hubig:_syten_toolk,hubig17:_symmet_protec_tensor_networ,hubig_2015,
	McCulloch_2007}, we keep up to $20000$ SU(2) states, 
(typically equivalent to $\gtrsim 80000$ U(1) states). 
We calculate the ground-state energy, the spin structure factor and low-energy 
excitations for these clusters, yielding an estimate for 
the ground-state energy per site in the thermodynamic limit of $E_0/N_\text{sites} = -0.490(6)$.  
The study of finite size clusters is complemented by a high order NLCE calculation, which 
\emph{excludes} any scenario where $E_0/N_\text{sites}>-0.471$.

Our main finding is that the ground state of the larger (64-, 108- and 128-site) clusters we consider exhibits a breathing instability,
rendering up and down tetrahedra (cf. inset of
Fig.~\ref{fig:energy_comparison})
inequivalent: one tetrahedral sublattice exhibits a lower energy than the other.  Amusingly, our  estimate for the 
ground state energy is compatible with that of the original perturbation theory with a simple 
mean field solution of the resulting effective Hamiltonian, where the 
inversion symmetry was maximally broken at the very outset of the 
calculation \cite{harris_ordering_1991}.  

\emph{Model and methods.---}
We consider the  pyrochlore antiferromagnetic Heisenberg model with $S=1/2$:
\begin{align}
	H = J \sum_{\langle i, j \rangle} \vec{S}_i \cdot \vec{S_j},
	\label{eq:Hamiltonian}
\end{align}
where the spins sit on the sites $i,j$ of the 3D pyrochlore lattice and $\langle 
i, j\rangle$ denotes nearest neighbors. The lattice is 
a face centered cubic lattice with lattice vectors 
$\vec{a}_1=\frac{1}{2}(1,1,0)^T$, $\vec{a}_2=\frac{1}{2}(1,0,1)^T$, 
$\vec{a}_3=\frac{1}{2}(0,1,1)^T$ and a tetrahedral basis given by 
$\vec{b}_0=\vec{0}$, $\vec{b}_1=\frac{1}{2}\vec{a}_1$,  
$\vec{b}_2=\frac 1 2 \vec{a}_2$,  $\vec{b}_3=\frac{1}{2}\vec{a}_3$, such that 
each lattice point can be expressed by
$\vec{R}_{\alpha,n_1,n_2,n_3} = n_1 \vec{a}_1 +  n_2 \vec{a}_2 +  n_3 
\vec{a}_3 + \vec{b}_\alpha,$
with integer $n_1,n_2,n_3$ and $\alpha \in \{0,1,2,3\}$. 
The model is obviously SU(2) symmetric.
Our DMRG calculations are performed on finite size ($N=32, 48, 64, 108, 128$) 
clusters with periodic boundary conditions  (cf. Tab. \ref{tab:clusters} of 
Appendix).

We apply the one- and two-site variants of SU(2) DMRG to reach high bond 
dimensions necessary to obtain reliable results in our three-dimensional 
clusters. Since DMRG requires a one-dimensional topology, we impose a one-dimensional ``snake'' path 
on the three-dimensional lattice, which defines the variational manifold. We use 
fully periodic clusters to reduce boundary effects and confirm that  
using a snake path which minimizes the bandwidth of the connectivity matrix 
improves convergence 
\cite{schollwock_review_2005,ummethum_numerics_2013,schafer_pyrochlore_2020}.

For small bond dimensions ($\chi \lesssim 2000$) we use the two-site version 
of the DMRG, and switch to the one-site variant to optimize the wave 
function for larger $\chi$. Since the truncation error is not well defined in the one-site 
variant case (due to the subspace expansion \cite{hubig_2015}), we use the reliable two-site variance estimation to extrapolate 
towards the error-free case \cite{hubig_prb_2018}, because calculation of the 
full variance would be impractical due to its cost. 

It turns out that even 
the calculation of the two-site variance becomes too costly for clusters with 
more than $\sim 100$ sites and bond dimensions $\gtrsim 8000$. In certain cases, we revert to the usage of the two-site DMRG and extrapolate as a function of the truncation error (cf. 108-site cluster).

\begin{figure}[t]
    \centering
\includegraphics[width=\columnwidth]{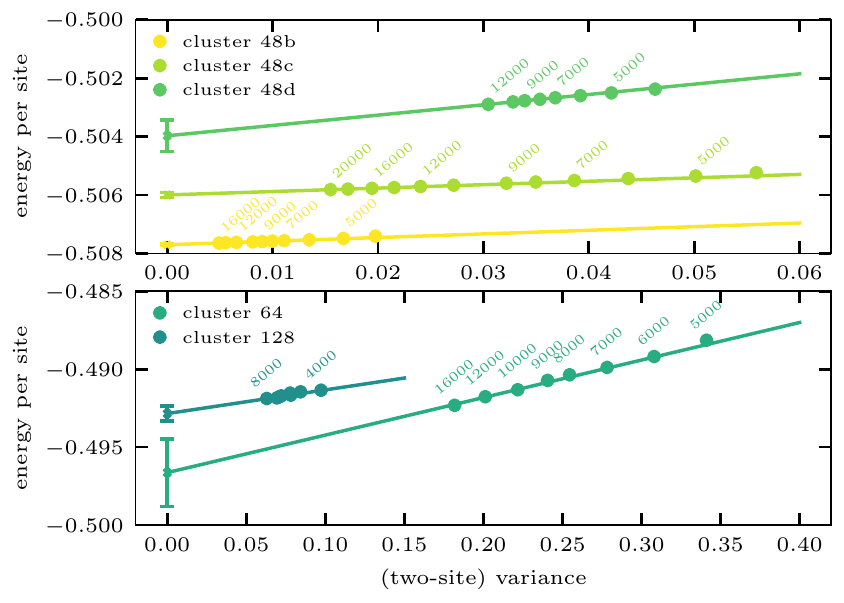}
\caption{Variational ground state energy estimates of the clusters 48b, 48c, 
48d (top) and 64,128 (bottom) for different bond SU(2) bond dimensions $\chi$ 
(indicated by the labels) as a function of the two-site variance. Solid lines 
correspond to linear extrapolations to the error-free limit, corresponding to 
infinite bond dimension and zero variance. We  estimate the systematic 
extrapolation error as the half distance between the last point and the 
extrapolated value.}
        \label{fig:energy_extrapolation}
\end{figure}

\emph{Ground-state energy.---}
Using DMRG, we calculate the variational ground-state energy of finite clusters 
with high accuracy.
By systematically increasing the bond dimension $\chi$, we enlarge the 
variational manifold in a controlled way, such that we can extrapolate, $\chi\to\infty$, to the exact limit using a linear extrapolation as a function of the two-site variance (cf. Fig. 
\ref{fig:energy_extrapolation}). We use an estimate of the systematic extrapolation error given by half the distance between the extrapolated value and the last DMRG point.

Fig.~\ref{fig:energy_comparison} shows the extrapolated energies per lattice site of all finite clusters we considered in comparison with the available predicted ground-state energies in the literature. 
Our results show a monotonic growth of the ground-state energy as the number of sites is increased. 

The periodic clusters we consider have either the full cubic (32, 108) or an 
increased or reduced (48a, 48b, 48c, 48d, 64, 128) symmetry of the pyrochlore lattice and 
represent the bulk due to the absence of a surface.
The energies per site of different clusters as a function of inverse cluster size admit 
a fit to a quadratic polynomial, which we use to obtain an extrapolation to the thermodynamic limit. In order to get an estimate of the extrapolation error, we use Gaussian resampling, using the systematic DMRG error-bars as standard deviation. This yields our best estimate for the ground-state energy of $E_0/N_\text{sites} = -0.490(6)$.
In this fit we considered only the cluster 48d among the 48-site clusters, which appears 
to be consistent with the other clusters, while other 48-site clusters have lower ground state energies.

Our extrapolated ($\chi\to\infty$) cluster energies and gaps are summarized in Table~\ref{table:0}. 
While the singlet gaps in the most symmetric 
clusters (32, 48d) are very small, the triplet gaps are sizable and roughly an 
order of magnitude larger. Since the 48d cluster does not obey all lattice 
symmetries, a reliable extrapolation is not possible, but our results are compatible
with a scenario with a finite triplet gap, in which case
all low energy excitations would be in the singlet sector as claimed in Refs. 
\cite{Berg_subcontractor_2003, tsunetsugu_prb_2001}.

Our finite temperature NLCE \cite{rigol_nlce_kagome_2006, rigol_nlce_square_2007, rigol_nlce_kagome_square_tri_2007, 
	khatami_nlce_checkerboard_2011, khatami_nlce_clinoatacamite_2011, 
	khatami_nlce_pinwheel_kagome_2011,  khatami_nlce_optical_lattic_2011, khatami_nlce_hubbard_2011, applegate_nlce_pyrochlore_exp_comp_2012, 
	singh_nlce_tetra_pyrochlore_2012, tang_nlce_2013, hayre_nlce_pyrochlore_2013, 
	jaubert_nlce_pyrochlore_exp_comp_2015, benton_nlce_pyrochlore_qsl_2018, 
	benton_nlce_pyrochlore_2018, pardini_nlce_pyrochlore_qsl_2019,  
	schafer_pyrochlore_2020}
provides a complementary perspective. 
We have carried out this expansion in 
entire tetrahedra up to eighth order (cf. \cite{schafer_pyrochlore_2020} for 
details, as well as Appendix \ref{sec:nlce}),  obtaining 
convergence for the energy per site in the thermodynamic limit as 
a function of temperature for temperatures $T\gtrsim 0.2$. Since the energy is a 
monotonic function of temperature, the converged part of $E(T)$ (cf. Fig. 
\ref{fig:nlce_00}) provides an upper bound for the ground-state energy 
$E_{\text{nlce}} \approx -0.471J$, which is consistent with the DMRG data and extrapolation.
One can furthermore polynomially 
extrapolate the finite temperature NLCE energies to zero 
temperature (assuming an analytic behavior at low temperatures), see Fig. \ref{fig:nlce_00}, and obtain
$-0.495(15)$, which agrees remarkably well with the DMRG extrapolation and lies  
within its error bar, serving as a further corroboration of the DMRG energy.
In light of these results we can confidently exclude a ground-state 
energy per site larger than $-0.47J$. 

\begin{table}[!htb]
\centering
\begin{tabular}{@{}c|@{\hspace{4mm}}l@{\hspace{4mm}}l@{\hspace{4mm}}l@{\hspace{ 4mm}}}
\hline\hline
Cluster & GS energy & Singlet gap & Triplet gap  
\tabularnewline\hline
32& $-0.5168$ & 0.0318 & 0.6872 \tabularnewline
48a & $-0.5161$ & 0.2166(4) & 0.6709(4) \tabularnewline
48b & $-0.5077$ & 0.027(2)& 0.554(2) \tabularnewline
48c & $-0.5060(1)$ & 0.053(7) & 0.42(2) \tabularnewline
48d & $-0.5040(5)$ & 0.06(3) &0.36(3) \tabularnewline
64& $-0.4972(25)$& --- & --- \tabularnewline
108& $-0.4935(50)$& --- & --- \tabularnewline
128& $-0.4928(10)$& --- & --- \tabularnewline\hline\hline
\end{tabular}
\caption{Ground-state energies per site and gaps within the $S_\text{tot}=0$ 
sector (singlet gap) as well as to the $S_\text{tot}=1$ sector (triplet gap) if 
available.}
\label{table:0} 
\end{table}

\emph{Ground-state symmetry-breaking.---}
To investigate the properties of the ground state in more detail,
we calculate the total spin, and hence total energy, of up and down tetrahedra separately. This reveals an inequivalence of up and down tetrahedra (cf. Fig. \ref{fig:tetra_extrap} in the Appendix), suggesting a breaking of the inversion symmetry of the lattice. 
\begin{figure}
	\centering
\includegraphics[width=\columnwidth]{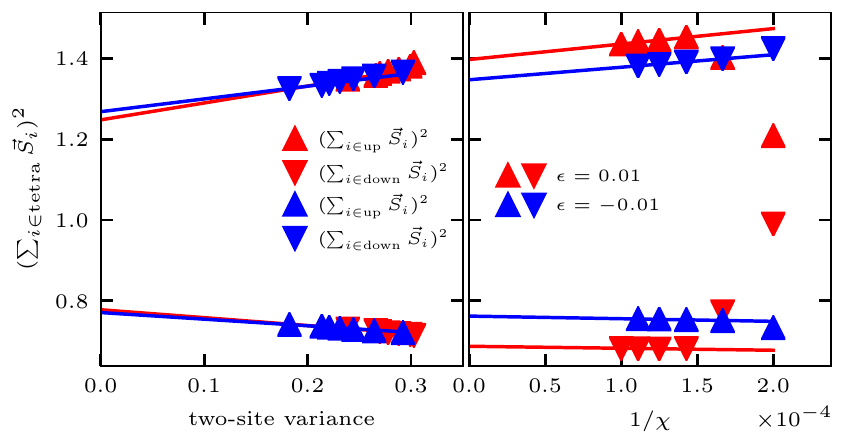}
	\caption{Extrapolation of tetrahedron spins for an explicit breaking of 
		lattice inversion symmetry, similarly to a ``pinning'' coupling, 
		for the 64 (left) and 108 (right) site clusters. The whole 
		Hamiltonian (\ref{eq:Hamiltonian}) is written as 
		$H=(1-\epsilon)H_\text{up}+(1+\epsilon)H_\text{down}$, where the $H_\text{up}$ 
		and $H_\text{down}$ parts contain the terms for the up and down 
		tetrahedra, 		respectively.  \label{fig:symbreak}}
\end{figure}
In our DMRG calculations, the snake path does not fully respect the
symmetry between up and down tetrahedra, so we need to
verify that this symmetry breaking is intrinsic, and not
due to a preference imposed by the snake path. We therefore introduce
a small symmetry breaking `breathing' perturbation, where we modify
the couplings of up and down tetrahedra to be $J=1\pm \epsilon$, equivalent to
the standard technique of including pinning fields.

Fig. \ref{fig:symbreak} shows the results for the total spin of up and
down tetrahedra for opposite signs of the breathing perturbation in
the 64 (108) site clusters as a function of the two-site variance (inverse bond dimension), admitting a linear extrapolation towards $\chi\to\infty$. 
The results reveal a clear selection of states
with opposite symmetry breaking, as required for spontaneous
symmetry-breaking. The order parameters for the larger, 108-site,
cluster are slightly different for the two opposite pinning fields
(Fig. \ref{fig:symbreak}, right panel), but that difference is much
smaller than the extrapolated order parameter which differs only little between the two clusters. It is of course always
possible in principle that the symmetry breaking vanishes when yet
larger clusters are considered. Given the scaling of the computational
effort with system size, the study of much larger clusters with the
present method is, however, out of reach. In Appendix \ref{sec:symbreak} we provide further evidence that the two symmetry-breaking states
converge to the same energy after the pinning field is removed.

\begin{figure}[t]
	\includegraphics[width=0.45\columnwidth]{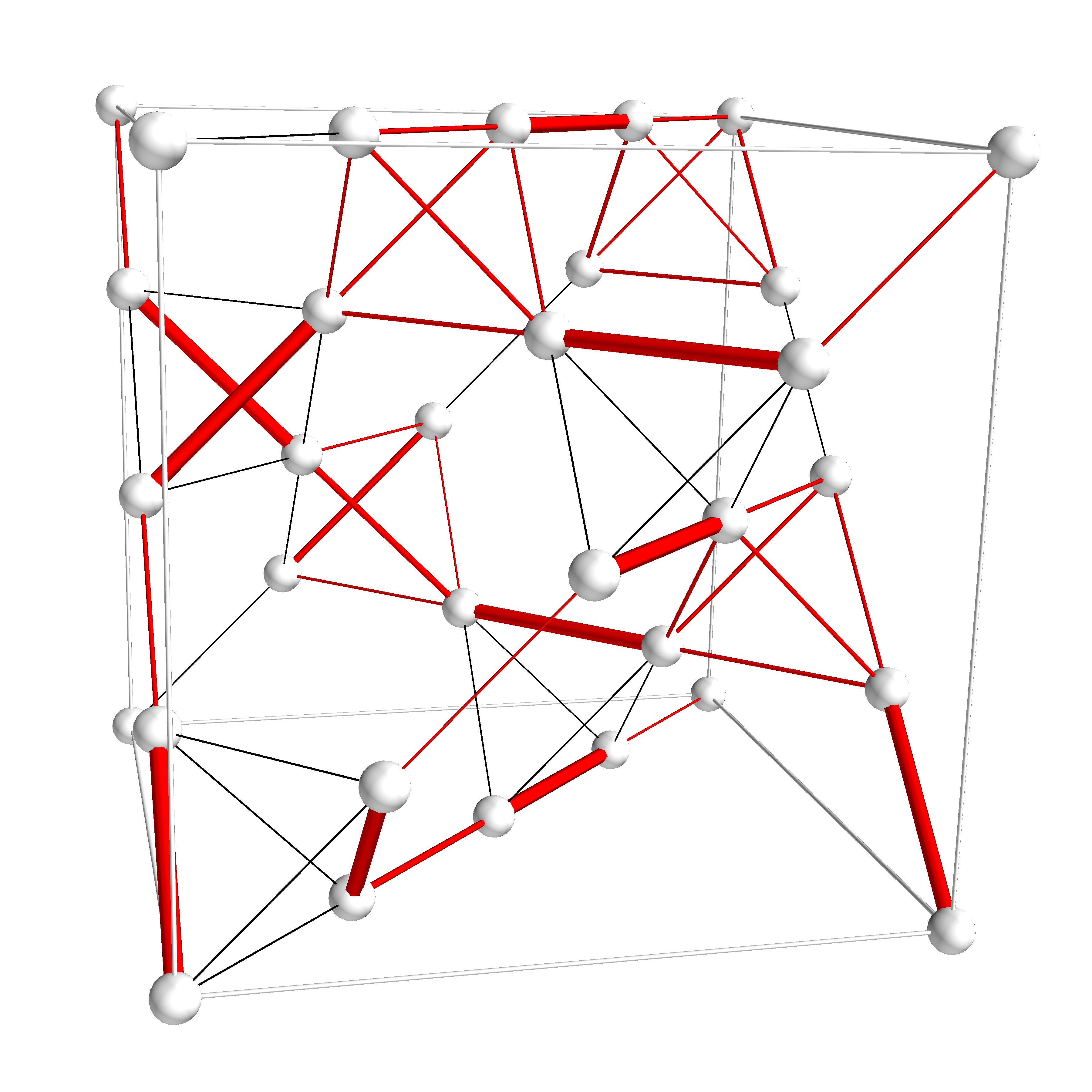} 
	\includegraphics[width=0.45\columnwidth]{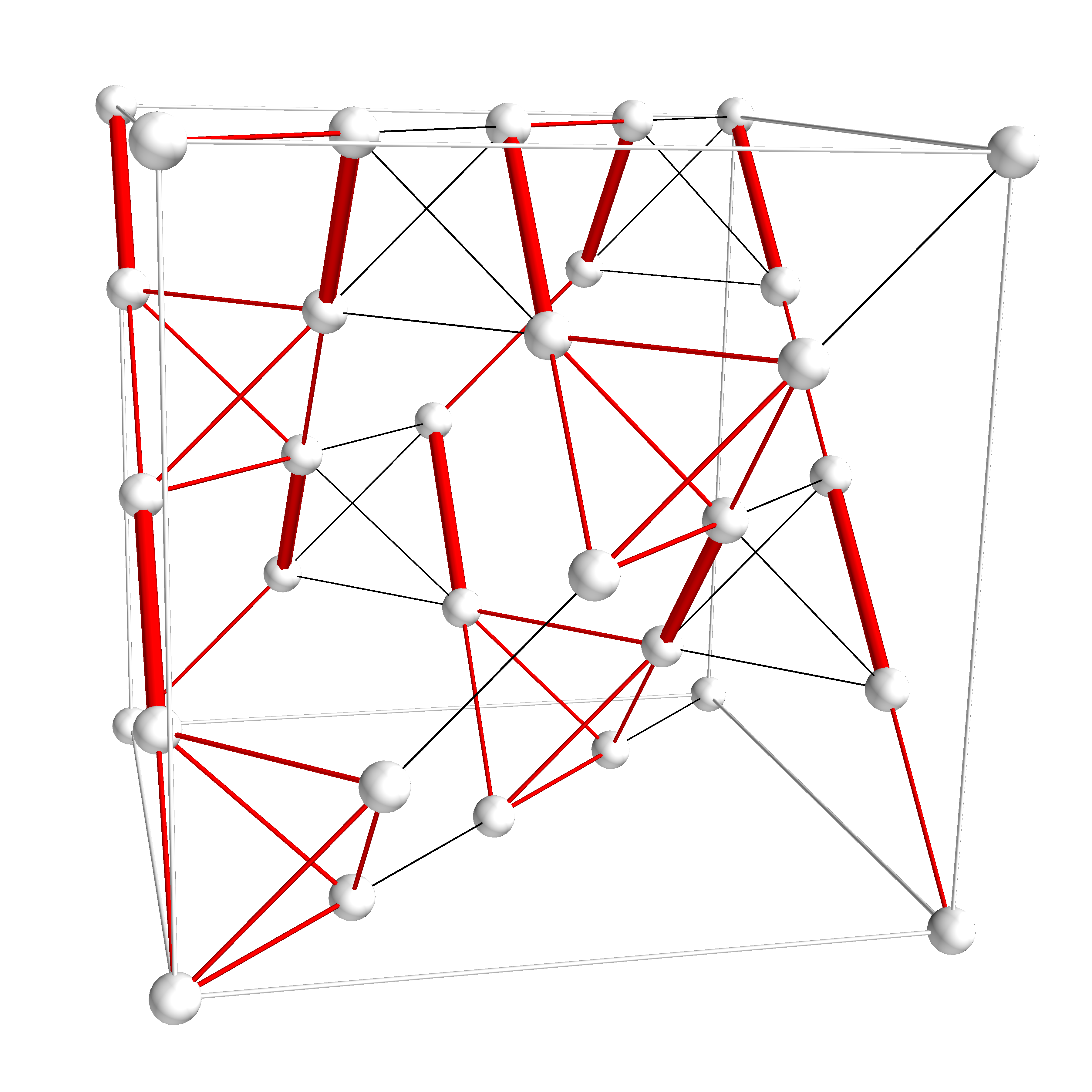} 
	\caption{Real space spin correlation $C_{ij}$ in the ground state ($S_z=0$) for $N=64$ (left) and $N=128$ (right) shown in the cubic unit cell. The thickness of the red bonds corresponds to magnitude of the correlation between neighboring sites. The black lines indicate bonds between sites with negligible correlations.} \label{fig:realspace_correlation}
\end{figure}

We next consider nearest neighbor spin correlations 
of the best (lowest-energy) wave functions $\ket{\psi_0}$ obtained in DMRG. 
For each pair of adjacent sites $(i, j)$, we calculate the correlation function 
$C_{ij}=\bra{\psi_0} \vec{S}_i \cdot \vec{S}_j \ket{\psi_0}$. We plot the result 
for the clusters $64$ and $128$ in Fig. \ref{fig:realspace_correlation} (truncated 
to the cubic unit cell for ease of visualization), with the tube thickness proportional to the strength of the spin correlations. 

The correlation pattern reveals that one sublattice (say, `up') of
tetrahedra contains more strongly correlated bonds than the other. These are found on opposite edges of `up' tetrahedra.
We note that the details of this pattern still depend strongly on the cluster geometry and we get opposite choices of correlated bonds in the two clusters, presumably due to different symmetry broken states picked by the different `snake' paths in the two clusters. Moreover, the periodic boundary conditions impact the performance of the DMRG calculation. In particular, finite-sized clusters with periodic boundary conditions comprise winding loops which may be as short as, or even shorter, than the `physical' loops in the bulk, whose minimal length is the circumference  (6) of a hexagon. Resonances along both loop types will therefore compete. The minimal length of winding loops for $N=108$ is $6$ while  it is $8$ for $N=128$. Indeed,  we observe considerably better convergence for the latter, inducing a smaller error, see Fig. \ref{fig:energy_comparison}. The shortest periodic loop of each cluster is shown in Tab. \ref{tab:clusters}.

\begin{figure}
	\centering
	\includegraphics[width=\columnwidth]{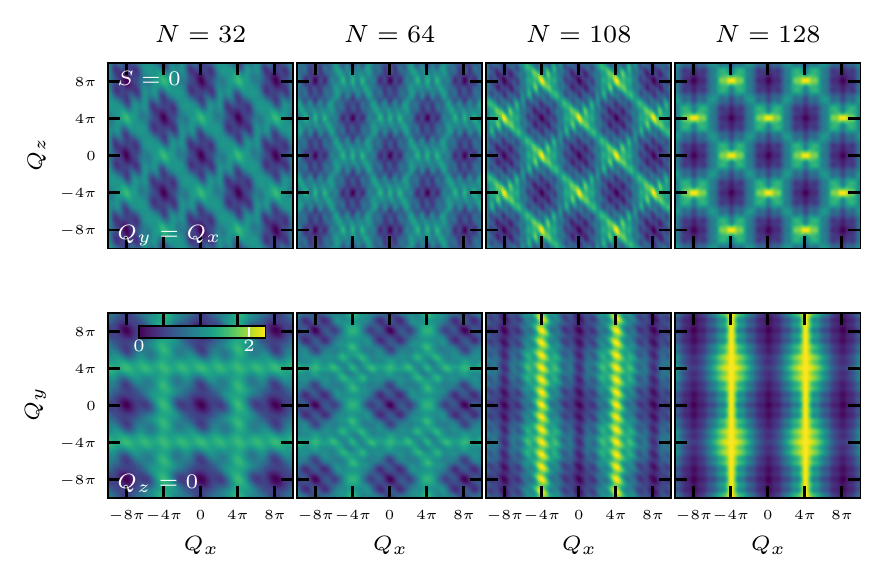}
	\caption{Static spin structure factor for different clusters for two 
cuts ($Q_x=Q_y$ (top) and $Qz=0$ (bottom)) through momentum space. The 
corresponding maximal bond dimensions for the $32,64,108$ and 128-site clusters are 
20000,16000, 16000 and 12000 respectively.}
	\label{fig:structure_factor}
\end{figure}

\emph{Ground-state structure factor.---} The static spin structure factor for different clusters, accessible in 
neutron scattering experiments, is obtained from the Fourier transform of 
the spin correlations (factor $4/3$ from normalization 
$1/(S(S+1))$ for spin $S=1/2$):
\begin{equation}
    S(\vec{Q})= \frac{4}{3N} \sum_{i j} \langle \vec{S}_i\cdot 
\vec{S}_j\rangle_c \cos\left[\vec{Q}\cdot 
\left(\vec{R}_i-\vec{R}_j\right)\right],
\end{equation}
where $\vec{R}_i$ denote the real-space coordinates of sites and the index $c$ 
denotes the connected part of the correlation matrix.
The results for two cuts ($Q_x=Q_y$ (top) and $Q_z=0$ (bottom)) in the three-dimensional momentum space are shown in 
Fig.~\ref{fig:structure_factor}.

One can readily recognize the bow-tie patterns, the hallmark of
pyrochlore magnets \cite{harris_ordering_1991,CanalsLacroix_prl,Isakov_dipolar_prl,
  muller_thermodynamics_2019,iqbal_quantum_2019,plumb_continuum_2019,
  schafer_pyrochlore_2020}.  Note that the 32- and 108-site clusters
have full cubic symmetry, while the 64-site cluster does not, hence
the structure factors looks slightly different in that case.  The
results for the spin structure factor and the absence of sharp Bragg
peaks confirm that there is no long range magnetic ordering. The observed pattern for the $Q_x=Q_y$ cuts is very close to what is found at
finite temperature in the regime $T\lesssim 1$
\cite{schafer_pyrochlore_2020}, on the other hand the $Q_z=0$ cuts exhibit a drastic change in the 108- and 128-site clusters reflecting the symmetry breaking. While the pinch points sharpen with
increasing system size (and therefore momentum resolution), we are
unable to extrapolate their width reliably to the thermodynamic limit
to extract a correlation length. 
Note that for the largest clusters, apparent lines in the spin structure factor in the $Q_x$-$Q_y$ plane become discernible, Fig. \ref{fig:realspace_correlation}, raising the possibility of at least short-range spin correlations with spatial anisotropy. A more detailed search for such symmetry breaking is clearly warranted.

\emph{Concluding discussion.---} Our DMRG study has found the ground
state of the $SU(2)$ symmetric $S=\frac 1 2$ Heisenberg
antiferromagnet to discard lattice inversion symmetry in favour of a
`breathing' pattern of strong (weak) sublattices of up (down)
tetrahedra. We extrapolate the energy per lattice site to
$-0.490(6)$. The possibility of such {\it spontaneous} symmetry
breaking has been a central question for this class of magnets, as
several studies have used an {\it explicit} such symmetry breaking as
a starting point of various perturbative
schemes~\cite{harris_ordering_1991,tsunetsugu_prb_2001,Berg_subcontractor_2003,
MoessnerGoerbig_prb_2006}. As the
restoration of an explicitly broken symmetry in a perturbative scheme
is generically not to be expected, a nonvanishing order
parameter does not per se indicate spontaneous symmetry breaking. 

Our results are thus important in that they provide largely unbiased
evidence for the existence of this spontaneous symmetry-breaking,
subject only to finite-size effects which are much reduced in
comparison to previous studies. This also indicates that one
of the prime Heisenberg quantum spin liquid candidates in three
dimensions in fact exhibits at least one form of symmetry breaking.

In closing, we note that our extrapolated ground-state energy lies
close to the estimate obtained in the pioneering work by Harris
\emph{et al.}  [\onlinecite{harris_ordering_1991}], in the
abovementioned scheme of coupling the up tetrahedra perturbatively
through the bonds of the down tetrahedra. These authors also found a
long-range dimer ordering (cf. also \cite{tsunetsugu_prb_2001})
compatible with the correlation pattern we observe in our calculations
shown in Fig. \ref{fig:realspace_correlation}. This first, simple and
quite uncontrolled, approach to this difficult problem thus may turn
out to have been already quite close to what will eventually be
established as the final answer.
\begin{acknowledgments}
	We thank Owen Benton, Ludovic Jaubert, Paul McClarty, Jeffrey Rau, Johannes Richter, Oleg Derzhko, 
	Masafumi Udagawa and Karlo Penc for very helpful discussions.
	We acknowledge financial support from the Deutsche Forschungsgemeinschaft
	through SFB 1143 (Project-id 247310070) and cluster of excellence 
ct.qmat (EXC 2147, Project-id 390858490).
I.H.~was supported in part by the Hungarian National Research,   
Development   and   Innovation Office (NKFIH) through Grant No.~K120569 and 
the Hungarian  Quantum  Technology  National  Excellence Program  (Project  
No.~2017-1.2.1-NKP-2017-00001). Some of the data presented here was produced 
using the \textsc{SyTen} toolkit, originally created by Claudius 
Hubig.\cite{hubig:_syten_toolk,hubig17:_symmet_protec_tensor_networ}
\end{acknowledgments}

\bibliography{pyrochlore}
\clearpage
\appendix

\section{Inversion-symmetry breaking}
\label{sec:symbreak}

\begin{figure}[h]
	\centering
	\includegraphics[width=\columnwidth]{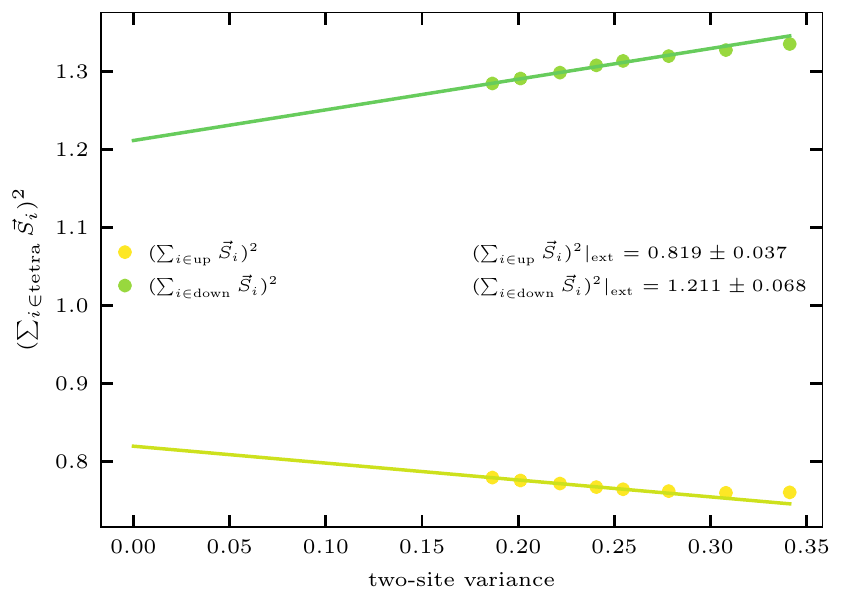}
	\caption{The extrapolation of the total spin squared as a function of the 
		two-site variance for the 64-site cluster.}
	\label{fig:tetra_extrap}
\end{figure}

\begin{figure}[h]
	\centering
	\includegraphics[width=\columnwidth]{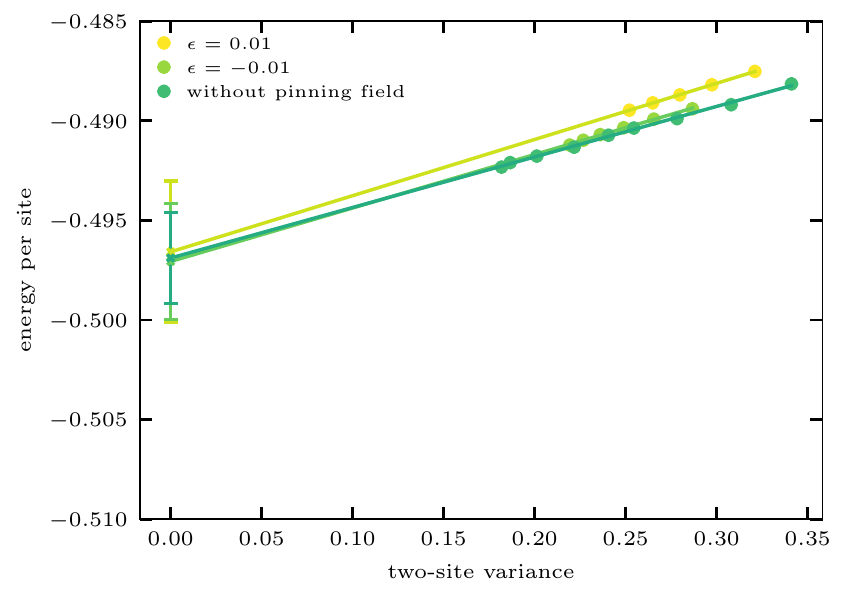}
	\caption{The extrapolation of the energies after the pinning field is switched off together with the one without the pinning field for the 64-site cluster.}
	\label{fig:energy_field_off}
\end{figure}

The real space dimer correlation pattern shown in Fig. \ref{fig:realspace_correlation} suggests that the lattice inversion symmetry is broken in the ground state. In order to scrutinize this finding, we analyze the square of the total spin (morally the tetrahedron energy) of up and down tetrahedra in the lattice separately. Fig. \ref{fig:tetra_extrap} shows the extrapolation to the exact limit for the 64 site cluster. The extrapolation clearly suggests an imbalance between up and down tetrahedra, and confirms the finding from the real space dimer correlations. This is further corroborated by a high susceptibility towards inversion symmetry breaking perturbations, as discussed in the main text.
In order to determine whether the applied pinning, $\epsilon=0.01$ is sufficiently small, we apply  the following DMRG procedure. For a given ordering of the sites, the symmetry-breaking state involving the least entanglement, is preferred by DMRG. Our strategy is as follows: we perform DMRG calculations up to a certain bond dimension (with the pinning appied), until the states are stabilized. Then we switch off the pinning ($\epsilon\to0$) and perform further sweeping and increase the bond dimension. If the symmetry breaking is intrinsic, both of these energies should agree with each other, and with the one without pinning. We carried out this test for the 64-site cluster and find that if the pinning is removed too early (for example  with 3000 states) DMRG converges back to the state preferred by the snake. However, the state becomes stable at bond dimension 5000 and remains so for the further increase of the bond dimension and sweeping. Obviously, this problem is not present if the pinning  as well as the snake prefer the same state. Considering the energies, a smooth linear extrapolation is possible (Fig.~\ref{fig:energy_field_off}) and each of them results in the same energy lying well within the error bars.

\section{Other lattice symmetries}
We further investigated various symmetries with respect to our ground state correlations shown in Fig. \ref{fig:realspace_correlation} for the clusters with $N=64$ and $N=128$. As mentioned in the main text, resonant loops across  periodic boundary conditions compete with the loops in the bulk. 
The cluster with $N=108$ exhibits loops across the boundary of length $6$ which compete with bulk hexagonal loops. Therefore, the correlation pattern of the $N=108$ appears defective in some regions if compared to the pattern obtained for $N=128$ with loops across the boundary of length 8. For the $N=64$ cluster, also boundary loops of length $6$ appear. The last column of table \ref{tab:clusters} lists the length of the shortest boundary loops in all clusters we investigated.

To analyze the symmetries of the observed correlation pattern, we start by coarse graining it.
There are three different types of bonds. First, we find six uniformly weakly coupled bonds ($A$) on one type of the tetrahedra. Second, the inverted tetrahedra exhibit two types of bonds, two strong dimers ($B$) and four vanishing bonds ($C$).  The average correlation strengths of each bond type are listed in Table \ref{tab:averaged_bonds} for $N=64$ and $N=128$. The simplified nearest neighbor correlation pattern can then be viewed as a graph with edges labeled by $A,B$, or $C$.


Each lattice symmetry is a permutation $\pi \in S_N$ of nodes in this graph. The type $X$ of an edge $(i,j)$ is preserved under the symmetry transformation if $(\pi(i),\pi(j))$ is of the same type $X$. Applying all lattice symmetries to the labeled graphs for our largest clusters, $N=64$ and $N=128$, we can count how many edges preserve their type under the symmetry and list the ratio the ratio $c/t$ of conserved edges $c$ over the number of total edges $t=3N$ in Table \ref{tab:symmtries} for all lattice symmetries.
\begin{table}[h]
    \centering
    \begin{tabular}{l@{\hspace{8mm}}c@{\hspace{8mm}}c@{\hspace{8mm}}c}
        \hline\hline
          & $A$ &  $B$ & $C$ \\
        \hline
        64 &  $-0.14339314$ & $-0.53707564$ & $-0.00869636$ \\
        128 &  $-0.14484361$ & $-0.53011351$ & $-0.00968693$ \\
        \hline
    \end{tabular}
    \caption{Averaged correlation strength of three types of $A$, $B$ and $C$ observed in the real space correlations in Fig. \ref{fig:realspace_correlation} starting from the simplified picture. $12$ of $192$ bonds for the $N=64$ and none of the $384$ bonds for $N=128$ violate the simplified picture.}
    \label{tab:averaged_bonds}
\end{table}

\begin{table}[h]
    \centering
    \begin{tabular}{l@{\hspace{8mm}}c@{\hspace{8mm}}c@{\hspace{8mm}}c}
        \hline\hline
          & $64$  & $128$ \\
        \hline
        $3^+_{(x,x,x)}$ & $136 / 192$ & $256 / 384$ \\
        $3^-_{(x,x,x)}$ & $136 / 192$ & $256 / 384$ \\
        $2_{(0,-y,y)}$ & $4/ 192$ & $0 / 384$ \\
        $2_{(-x,0,x)}$ & $12 / 192$ & $0 / 384$ \\
        $2_{(x,-x,0)}$ & $4 / 192$ & $0 / 384$ \\
        $I_{(0,0,0)}$ & $10 / 192$ & $0 / 384$ \\
        $3^+_{(x,x,x)}I_{(0,0,0)}$ & $10 / 192$ & $0 / 384$ \\
        $3^-_{(x,x,x)}I_{(0,0,0)}$ & $10 / 192$ & $0 / 384$ \\
        $m_{(x,y,y)}$ & $152 / 192$ & $384 / 384$ \\
        $m_{(x,y,x)}$ & $144 / 192$ & $256 / 384$ \\
        $m_{(x,x,y)}$ & $136 / 192$ & $256 / 384$ \\\hline
        $T_{(1,0,0)}$ & $136 / 192$ & $384 / 384$ \\
        $T_{(0,1,0)}$ & $136 / 192$ & $384 / 384$ \\
        $T_{(0,0,1)}$ & $132 / 192$ & $384 / 384$ \\
        $T_{(1,1,0)}$ & $132 / 192$ & $384 / 384$ \\
        $T_{(1,0,1)}$ & $136 / 192$ & $384 / 384$ \\
        $T_{(0,1,1)}$ & $136 / 192$ & $384 / 384$ \\
        \hline
    \end{tabular}
    \caption{Ratio of the number of conserved edges over the number of total edges $3N$ for different symmetry operations. $k^\pm_{(x,y,z)}$ describes a rotation by $\pm 2\pi/k$ around the axis $(x,y,z)$, $I_{(0,0,0)}$ is the inversion around the center and $m_{(x,y,z)}$ is a reflection. The second part of the table considers translations of the form $T_{(a,b,c)}$ which describes a shift by $a\vec{a}_1 + b\vec{a}_2 + c\vec{a}_3$.}
    \label{tab:symmtries}
\end{table}
From this systematic symmetry analysis, it is clear that the correlation pattern is 
fully symmetric for $N=128$ under all fcc translations $T_{(a,b,c)}$, which means that each tetrahedron shows the same orientation of strong dimers. This leads to quasi decoupled planes of weakly coupled tetrahedra connected by strong dimers in this cluster. 

On the other hand, all symmetry operations 
 matching one type of tetrahedra to another (in particular inversion) are robustly broken, since in both $N=64$ and $N=128$ clusters the number of type preserved bonds is essentially zero.  \\
 
The $N=64$ cluster confirms this picture. Due to the competition of length 6 boundary loops with bulk hexagonal looks, the correlation pattern is slightly defective if compared to $N=128$, which leads to slightly imperfect preservation of the pattern under symmetry operations. The orientation of the strong bonds B seem to be arbitrary.

\section{Finite-size clusters}
\label{sec:clusters}
We use the clusters 32, 48a, 48b, 48c, 48d, 64, 108, and 128 in our simulations, which are described by the cluster vectors $\vec{c_1}, \vec{c_2}, \vec{c_3}$. The performance of the DMRG calculation is  affected by loops winding across the periodic boundaries. The key element is the length of these winding loops compared to resonant loops within the bulk (predominantly hexagons). Therefore, Tab. \ref{tab:clusters} lists the length of the shortest loop connected via a periodic bond for each cluster.
\begin{table}[h]
    \centering
    \begin{tabular}{l@{\hspace{5mm}}c@{\hspace{5mm}}c@{\hspace{5mm}}c@{\hspace{5mm}}|c}
        \hline\hline
        cluster & $\vec{c}_1$ &  $\vec{c}_2$ & $\vec{c}_3$ & length  \\
        \hline
        32   &  $2 \vec{a}_1$ & $2 \vec{a}_2$ & $2\vec{a}_3$ & 4  \\
        48a   &  $(\frac{3}{2},\frac{1}{2},0)^T$ & $(0,1,1)^T$ & $(0,1,-1)^T$ & 4  \\
        48b   &  $(\frac{3}{2},\frac{1}{2},0)^T$ & $(0,\frac 1 2 ,\frac 3 2)^T$ & $(0,1,-1)^T$ & 4  \\
        48c   &  $(\frac{3}{2},1,\frac{1}{2})^T$ & $(0,1 ,-1)^T$ & $(1,-1,0)^T$ & 4  \\
        48d   &  $(1,1,1)^T$ & $(1 ,0,-1)^T$ & $(1,-1,0)^T$ & 4 \\
        64   &  $(1,1,1)^T$ & $(1 ,1,-1)^T$ & $(-1,1,1)^T$ & 6  \\
        108   &  $3 \vec{a}_1$ & $3 \vec{a}_2$ & $3\vec{a}_3$ & 6  \\
        128   &  $(2,0,0)^T$ & $(0 ,2,0)^T$ & $(0,0,2)^T$  & 8\\

%
        \hline
    \end{tabular}
    \caption{Cluster vectors $\vec{c}_1$,$\vec{c}_2$,$\vec{c}_3$ of the 8 clusters used in this work and the length of the shortest periodic loop. The clusters of size $32$ and $108$ respect all lattice symmetries.}
    \label{tab:clusters}
\end{table}

\section{Numerical linked cluster expansion}
\label{sec:nlce}
We apply a systematic high temperature series expansion to obtain an upper bound for the ground state energy of pyrochlore lattice in the thermodynamic limit. The numerical linked cluster expansion (NLCE) determines any extensive property $P$ (such as the energy) in the high temperature regime. It has been successfully applied to various geometries including frustrated systems like the kagome or pyrochlore lattice \cite{rigol_nlce_square_2007, 
rigol_nlce_kagome_square_tri_2007, tang_nlce_2013, rigol_nlce_kagome_2006, applegate_nlce_pyrochlore_exp_comp_2012, singh_nlce_tetra_pyrochlore_2012, khatami_nlce_optical_lattic_2011, khatami_nlce_hubbard_2011, 
khatami_nlce_checkerboard_2011, khatami_nlce_clinoatacamite_2011, 
khatami_nlce_pinwheel_kagome_2011,hayre_nlce_pyrochlore_2013, benton_nlce_pyrochlore_qsl_2018, 
pardini_nlce_pyrochlore_qsl_2019, jaubert_nlce_pyrochlore_exp_comp_2015, benton_nlce_pyrochlore_2018}.\par
A detailed description of the approach used here can be found in our previous work \cite{schafer_pyrochlore_2020}. It has been shown that an expansion based on \textit{tetrahedra} provides the most efficient approach, yielding reliably converged energy results down to temperatures $T\gtrsim 0.2$. Here, we include all clusters with full exact diagonalization consisting of up to 8 tetrahedra (i.e. up to $25$ spins $\frac{1}{2}$). These clusters include crucial loops of 6 and 8 spins.\par
Since the energy decreases monotonously with temperature, we are able use the converged part as an upper bound for the ground state energy ($E_{\text{nlce}} \approx -0.471J$) in the thermodynamic limit.
Assuming an analytic behavior we used the converged part in the finite temperature regime to predict the zero temperature ground state energy. Hence, we extrapolated the function using a quadratic polynomial:
\begin{align}
    E(T) = a + bT+ cT^2.
\end{align}
The range of the best fit is between the convergence limit at $T\approx 0.25$ and $T=0.5$, and we varied the range limits randomly to estimate the systematic error of the fit, yielding $E_{\text{extra}} \approx -0.495(15)$. 
\begin{figure}[h]
    \centering
\includegraphics[width=\columnwidth]{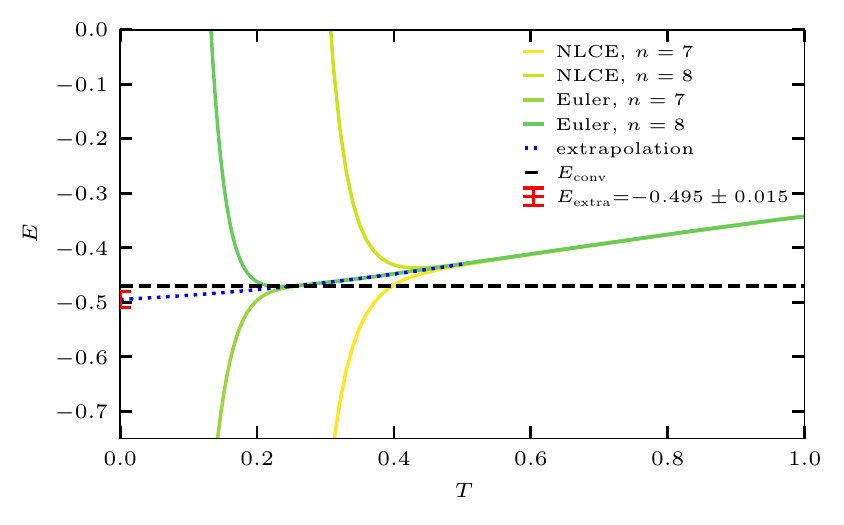}
    \caption{Energy per site with NLCE expansion up 8\textit{th} order in combination with the euler series acceleration with $k=3$\cite{schafer_pyrochlore_2020}. The energy in the thermodynamic limit is converged down to $T \approx 0.25$ in units of $J$ with a value of $E_{\text{nlce}} \approx -0.471J$. This can be used as an upper bound for the ground state energy. Additionally, we extrapolated the converged part with a simple quadratic \textit{ansatz} and received and an extrapolated ground state energy of $-0.495J$.}
        \label{fig:nlce_00}
\end{figure}
\end{document}